\documentclass[twocolumn,prl,showpacs,superscriptaddress,showpacs]{revtex4-1}
\usepackage{graphicx}
\usepackage{amssymb,amsmath}
\usepackage{bm}
\usepackage{dcolumn}
\usepackage{subfigure}
\usepackage{float}
\usepackage{url}
\usepackage{color}
\usepackage{txfonts}
\usepackage[driverfallback=dvipdfm]{hyperref}
\hypersetup{pdfpagemode=FullScreen,colorlinks=true,breaklinks,urlcolor=blue,linkcolor=blue,citecolor=blue}
\usepackage{natbib}

\begin{document}

\title{On the Equivalence between Spin and Charge Dynamics of the Fermi Hubbard Model}
\author{Hui Zhai}
\email{hzhai@tsinghua.edu.cn}
\affiliation{Institute for Advanced Study, Tsinghua University, Beijing, 100084, China}
\affiliation{Collaborative Innovation Center of Quantum Matter, Beijing, 100084, China}
\author{Ning Sun}
\affiliation{Institute for Advanced Study, Tsinghua University, Beijing, 100084, China}
\author{Jinlong Yu}
\affiliation{Institute for Advanced Study, Tsinghua University, Beijing, 100084, China}
\author{Pengfei Zhang}
\affiliation{Institute for Advanced Study, Tsinghua University, Beijing, 100084, China}

\date{\today }

\begin{abstract}

Utilizing the Fermi gas microscope, recently the MIT group has measured the spin transport of the Fermi Hubbard model starting from a spin-density-wave state, and the Princeton group has measured the charge transport of the Fermi Hubbard model starting from a charge-density-wave state. Motivated by these two experiments, we prove a theorem that shows under certain conditions, the spin and charge transports can be equivalent to each other. The proof makes use of the particle-hole transformation of the Fermi Hubbard model and a recently discovered symmetry protected dynamical symmetry. Our results can be directly verified in future cold atom experiment with the Fermi gas microscope.

\end{abstract}

\maketitle

Quantum gas microscope is one of the most significant developments in the cold atom physics during the past decade. It opens up a new avenue for studying strongly correlated physics, because it allows one not only to detect the system in situ with single-site resolution, but also to prepare an eigenstate of real space density operators, with which the non-equilibrium dynamics of strongly correlated systems can be studied. Recently, the MIT group and the Princeton group have prepared the Fermi Hubbard model (FHM) initially in a spin-density-wave state and a charge-density-wave state, respectively, and the subsequent spin or charge dynamics has been measured \cite{MIT, Princeton}. From these two measurements, they extracted the spin diffusion constant and the charge diffusion constant, respectively \cite{MIT, Princeton}. 

This article is to prove that, under certain conditions, the spin and the charge transport measurements can be equivalent to each other for the Fermi Hubbard model. To be specific, we first write down the FHM that these two groups have simulated by loading ultracold fermionic atoms in square optical lattices, that is
\begin{align}
\hat{H}=&-J\sum\limits_{\langle ij\rangle,\sigma}\hat{c}^\dag_{i\sigma}\hat{c}_{j\sigma}+U\sum\limits_{i}\left(\hat{n}_{i\uparrow}-\frac{1}{2}\right)\left(\hat{n}_{i\downarrow}-\frac{1}{2}\right), \label{FHMC}
\end{align} 
where $J$ is the hopping amplitude between two nearest neighbouring sites of the square lattice, and $U$ is the on-site interaction strength. Here the interaction term is written in a particle-hole symmetric form. Taking $J$ as the energy unit, the model is characterized by one single parameter $U$, together with two conserved quantities: $N_\uparrow+N_\downarrow-N_\text{s}$ ($N_\text{s}$ denotes the total number of sites), known as the doping from half filling; and $N_\uparrow-N_\downarrow$, known as the spin imbalance. 

\begin{figure}[t]
  \includegraphics[width=3.5in]{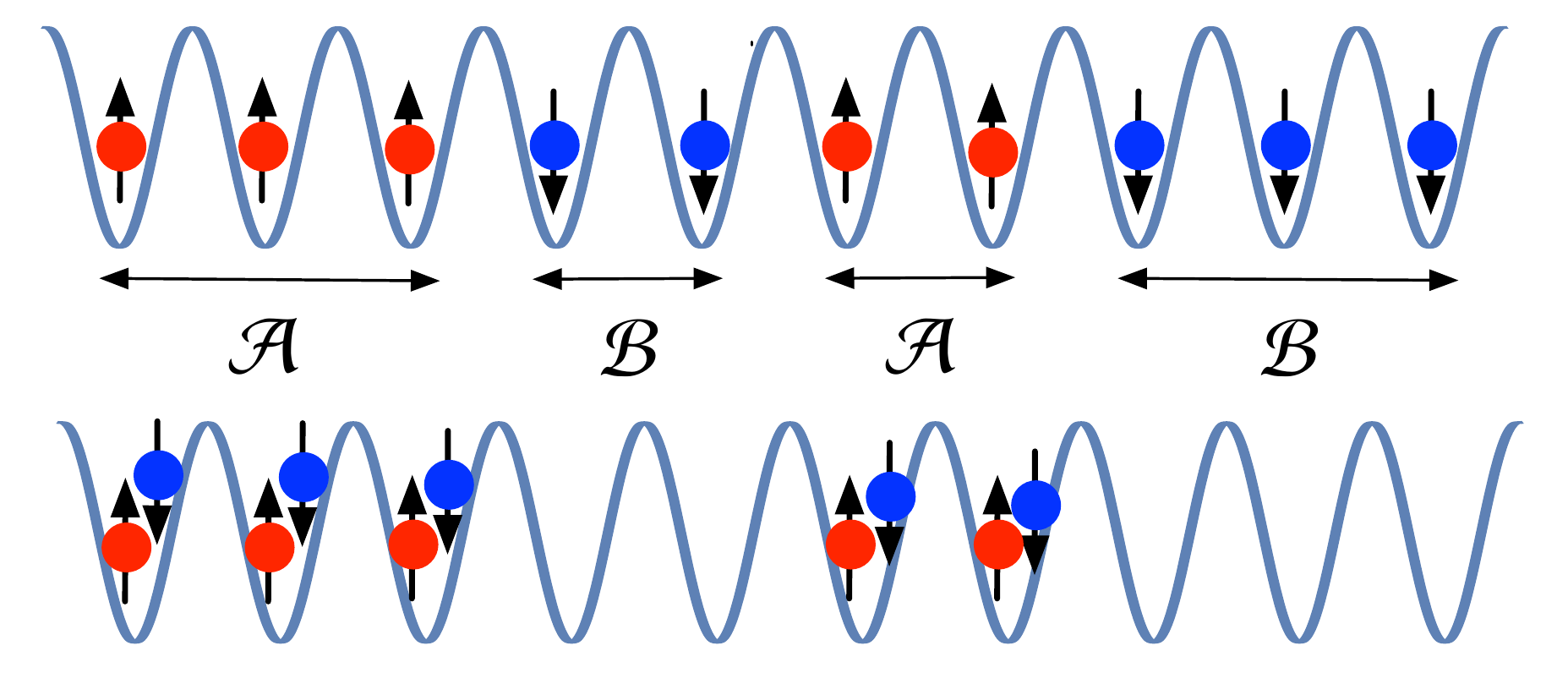}\\
  \caption{Schematics of a real space spin-density-wave state (upper panel) and the real space charge-density-wave state (lower panel) as the initial state for measuring spin and charge dynamics, respectively. }\label{schematic}
\end{figure}

First, let us start with a real space spin-density-wave state written as 
\begin{equation}
|\Psi\rangle_{\text{SDW}}=\prod\limits_{i\in \mathcal{A}}\hat{c}^\dag_{i\uparrow}\prod\limits_{j\in \mathcal{B}}\hat{c}^\dag_{j\downarrow}|0\rangle, \label{SDW}
\end{equation}
which is shown schematically in the upper panel of Fig. \ref{schematic}. Here, there is no constraint on the choices of region $\mathcal{A}$ and $\mathcal{B}$. Neither of them has to be single-connected or has equal size to the other. For instance, if one considers a $(\pi,\pi)$ anti-ferromagnetic state along $\hat{z}$ direction on a square lattice, then $\mathcal{A}$ denotes one sublattice and  $\mathcal{B}$ denotes the other. Or $\mathcal{A}$ denotes a group of domains where spins are polarized up, and $\mathcal{B}$ denotes the rest regions where spins are polarized down. This SDW state will then evolve under the FHM Hamiltonian and, at certain time $t$, one measures the local spin density along $\hat{z}$-direction as
\begin{equation}
S_i^z(t)={}_\text{SDW}\langle\Psi| e^{i\hat{H}t}(\hat{n}_{i\uparrow}-\hat{n}_{i\downarrow})e^{-i\hat{H}t}|\Psi\rangle_\text{SDW}. \label{spin}
\end{equation}

Similarly, we can write down an ideal version of charge-density-wave state that region $\mathcal{A}$ is doubly occupied while region $\mathcal{B}$ is empty, that is, 
\begin{equation}
|\Psi\rangle_{\text{CDW}}=\prod\limits_{i\in \mathcal{A}}\hat{c}^\dag_{i\uparrow}\hat{c}^\dag_{i\downarrow}|0\rangle. \label{CDW}
\end{equation}
This state is shown schematically in the lower panel of Fig. \ref{schematic}.
The evolution of this state is also governed by the FHM Hamiltonian, and at certain time $t$, one can measure the local total density, or its deviation from half filling, i.e., 
\begin{equation}
n_i(t)={}_\text{CDW}\langle\Psi| e^{i\hat{H}t}(\hat{n}_{i\uparrow}+\hat{n}_{i\downarrow}-1)e^{-i\hat{H}t}|\Psi\rangle_\text{CDW}. \label{charge}
\end{equation}

\textit{Theorem.} For the FHM on a square lattice, the measurement of the local spin density $S^z_i(t)$ defined by Eq. \ref{spin} with parameter $U_0$ and conserved quantities $N_\uparrow+N_\downarrow-N_\text{s}=x$ and $N_\uparrow-N_\downarrow=y$ always equals to the measurement of the local charge density $n_i(t)$ defined by Eq. \ref{charge} with the same $U_0$ and conserved quantities $N_\uparrow+N_\downarrow-N_\text{s}=y$ and $N_\uparrow-N_\downarrow=x$.

That is to say, the kind of charge and spin dynamics defined above are equivalent for the FHM of the same hopping and interaction parameters, with the doping and the spin imbalance quantities interchanging with each other. For instance, if one measures the spin dynamics of Eq. \ref{spin} for a half-filled FHM with spin imbalance, it is equivalent to measuring the charge dynamics of Eq. \ref{charge} for the same FHM with balanced spin population yet doped away from half filling. In particular, for a half-filled and spin-balanced FHM, the spin and charge dynamics defined above are always identical.  

\begin{figure}[t]
  \includegraphics[width=3.0in]{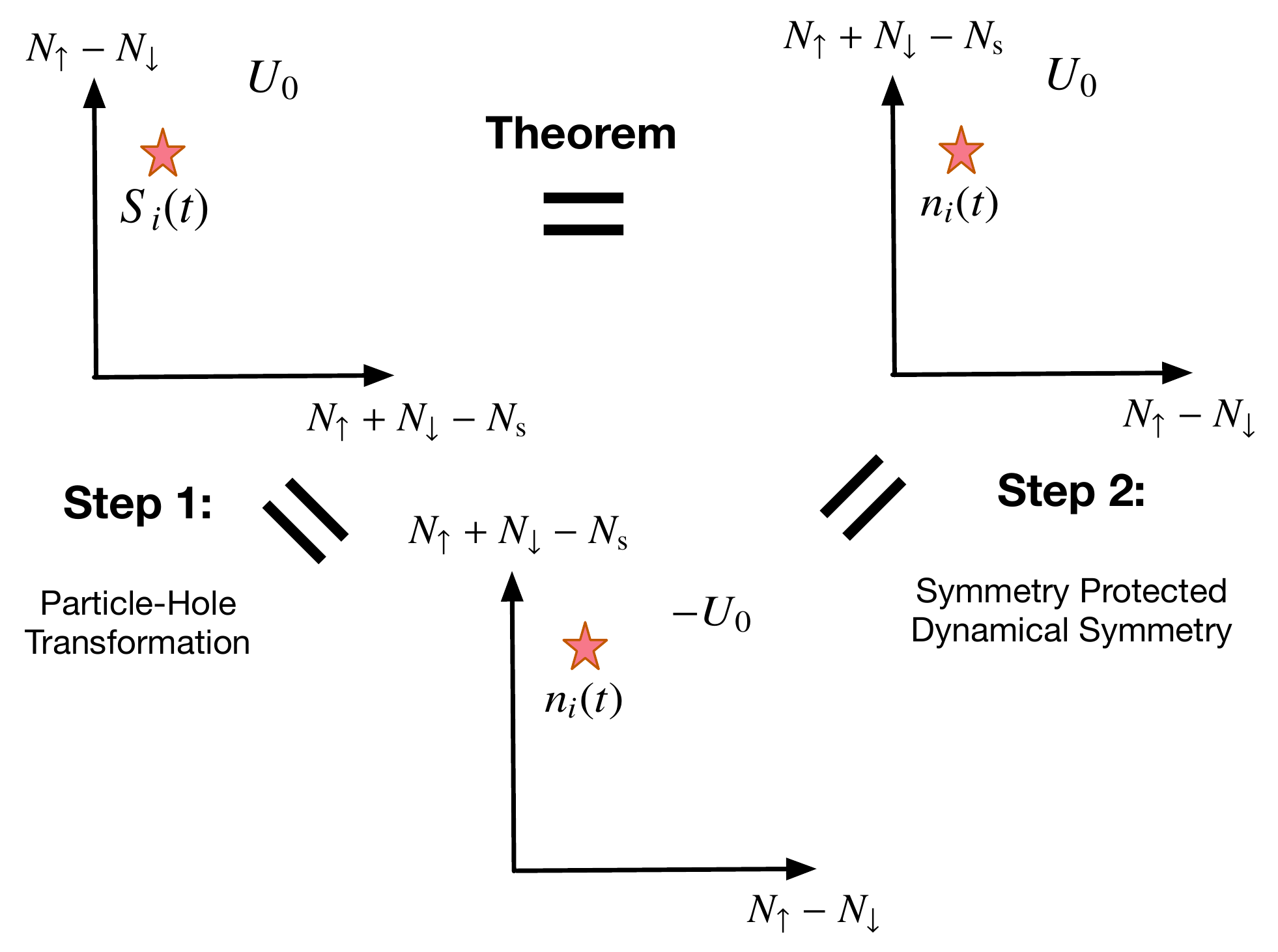}\\
  \caption{Schematic illustration of the theorem and two key steps for proving it. }\label{results}
\end{figure} 

The proof of this theorem follows from two steps. 

\textit{Step 1:} We consider a well-known particle-hole transformation $\mathcal{\hat{P}}$ defined as \cite{PH}
\begin{align}
&\hat{c}_{i \uparrow}\rightarrow \hat{c}_{i\uparrow}, \   \   \hat{c}^\dag_{i \uparrow}\rightarrow \hat{c}^\dag_{i\uparrow}\\
&\hat{c}_{i \downarrow}\rightarrow (-1)^{i_x+i_y} \hat{c}^\dag_{i\downarrow}, \   \   \hat{c}^\dag_{i \downarrow}\rightarrow (-1)^{i_x+i_y} \hat{c}_{i\downarrow},
\end{align}
where $i=(i_x, i_y)$ labels each site. This transformation leaves the spin-up field operators unchanged while makes a particle-hole transformation for the spin-down ones, accompanied by a sign change on one sublattice. This transformation does the following things. (i) It leaves the hopping term invariant, and inverts the sign of interaction term, i.e. $U \rightarrow -U$. (ii) Moreover, it interchanges the local spin density with the local particle density deviation from unity, 
\begin{equation}
(\hat{n}_{i\uparrow}-\hat{n}_{i\downarrow}) \longrightarrow (\hat{n}_{i\uparrow}+\hat{n}_{i\downarrow}-1),
\end{equation}
and it also interchanges the doping and spin imbalance of the system
\begin{align}
&(N_{\uparrow}-N_{\downarrow}) \longrightarrow (N_{\uparrow}+N_{\downarrow}-N_\text{s}),\\
&(N_{\uparrow}+N_{\downarrow}-N_\text{s})  \longrightarrow (N_{\uparrow}-N_{\downarrow}).
\end{align}
(iii) It also transforms the spin-density-wave state $|\Psi\rangle_\text{SDW}$ defined in Eq. \ref{SDW} to the charge-density-wave state $|\Psi\rangle_\text{CDW}$ defined in Eq. \ref{CDW}. 

As a result, the conclusion of the \textit{Step 1} is that the spin dynamics starting from the spin-density-wave state of Eq. \ref{SDW} with interaction parameter $U_0$ and conserved quantities $N_\uparrow+N_\downarrow-N_\text{s}=x$ and $N_\uparrow-N_\downarrow=y$ is equivalent to the charge dynamics starting from the charge-density-wave state of Eq. \ref{CDW} with interaction parameter $-U_0$ and conserved quantities $N_\uparrow+N_\downarrow-N_\text{s}=y$ and $N_\uparrow-N_\downarrow=x$. 

\textit{Step 2.} This step follows from another theorem we proved in Ref. \cite{SPDS}, which we termed as ``symmetry protected dynamical symmetry". It states as follows. 

Considering the Hamiltonian $\hat{H}=\hat{H}_0+\hat{V}$, here $\hat{H}_0$ is the single-particle hopping term and $\hat{V}$ the interaction term, if we can find an antiunitary operator $\mathcal{\hat{S}} = \mathcal{\hat{R}}\mathcal{\hat{W}}$, where $\mathcal{\hat{R}}$ is the (antiunitary) time-reversal operator and $\mathcal{\hat{W}}$ is a unitary operator that satisfy the following conditions: 

(i) $\mathcal{\hat{S}}$ anticommutes with $\hat{H}_0$ and commutes with $\hat{V}$, i.e. 
\begin{equation} \label{Eq:S_H}
  \{ \mathcal{\hat{S}},\hat{H}_0\}  = 0, \quad [\mathcal{\hat{S}},\hat{V}] = 0;
\end{equation}

(ii) The initial state $\left| {{\Psi }} \right\rangle $ only acquires a global phase factor under $\mathcal{\hat{S}}$, i.e. 
\begin{equation} \label{Eq:psi_0}
  \mathcal{\hat{S}}^{ - 1}\left| {{\Psi }} \right\rangle  = {e^{i\chi }}\left| {{\Psi }} \right\rangle;
\end{equation}

(iii) The measurement operator $\hat{O}$ is a Hermitian one that is even or odd under symmetry transformation $\mathcal{\hat{S}}$, i.e. 
\begin{equation} \label{Eq:SOS}
  \mathcal{\hat{S}}^{ - 1}\hat{O}\mathcal{\hat{S}} =  \pm \hat{O},
\end{equation}
then we can conclude
\begin{equation} \label{Eq:O_t_pmU}
  {\left\langle {O(t)} \right\rangle _ {+U} } =  \pm {\left\langle {O(t)} \right\rangle _ {-U} }.
\end{equation} 
Here $\left\langle {O(t)} \right\rangle _ {\pm U}$ denotes the expectation value of $\hat{O}$ under the time-dependent wave function $|\Psi(t)\rangle=e^{i\hat{H}t}|\Psi\rangle$ with interaction strength $\pm U$ in $\hat{H}$, respectively. 

Here we take $\mathcal{\hat{W}}$ as the bipartite lattice symmetry operation, defined as  
\begin{align}
\hat{c}_{i \sigma}\rightarrow (-1)^{i_x+i_y} \hat{c}_{i\sigma}, \   \   \hat{c}^\dag_{i \sigma}\rightarrow (-1)^{i_x+i_y} \hat{c}^\dag_{i\sigma}.
\end{align}
Unlike $\mathcal{\hat{P}}$, this transformation does not exchange particles and holes. Instead, it only introduces an extra minus sign on one sublattice for both two spin components. It is straightforward to check that with this choice of $\mathcal{\hat{W}}$ and with the initial state chosen as the charge-density-wave state defined in Eq. \ref{CDW}, conditions (i)-(iii) are satisfied. Moreover, the two conserved quantities $N_\uparrow+N_\downarrow-N_\text{s}$ and $N_\uparrow-N_\downarrow$ are both invariant under $\mathcal{\hat{W}}$. 

Thus, the conclusion of the \textit{Step 2} is that the charge dynamics starting from the charge-density-wave state of Eq. \ref{CDW} with interaction parameter $-U_0$ equals the charge dynamics from the same charge-density-wave state with interaction parameter $U_0$, with the same conserved quantities $N_\uparrow+N_\downarrow-N_\text{s}=y$ and $N_\uparrow-N_\downarrow=x$.

Combining the conclusions from the \textit{Step 1} and the \textit{Step 2}, the theorem is now proved. The theorem, as well as two steps of proof, is schematically shown in Fig. \ref{results}. From the proof, we can also see that the results can be more general in the sense that it does not depend on the specific choices of the initial state $|\Psi\rangle_\text{SDW}$ and $|\Psi\rangle_\text{CDW}$ introduced in Eq. \ref{SDW} and Eq. \ref{CDW}. We can measure the spin dynamics starting from $|\Psi\rangle_1$ as
\begin{equation}
S_i^z(t)={}_1\langle\Psi| e^{i\hat{H}t}(\hat{n}_{i\uparrow}-\hat{n}_{i\downarrow})e^{-i\hat{H}t}|\Psi\rangle_1, \label{spin2}
\end{equation}
and measure the charge dynamics starting from $|\Psi\rangle_2$ as 
\begin{equation}
n_i(t)={}_\text{2}\langle\Psi| e^{i\hat{H}t}(\hat{n}_{i\uparrow}+\hat{n}_{i\downarrow}-1)e^{-i\hat{H}t}|\Psi\rangle_\text{2}. \label{charge2}
\end{equation}
The theorem till holds as long as $|\Psi\rangle_1$ and $|\Psi\rangle_2$ satisfy the following two conditions:

(\textbf{1}) $|\Psi\rangle_1$ and $|\Psi\rangle_2$ are related to each other by the particle-hole transformation $\mathcal{\hat{P}}$;

(\textbf{2}) $|\Psi\rangle_2$ is invariant under $\mathcal{\hat{S}} = \mathcal{\hat{R}}\mathcal{\hat{W}}$ up to a phase, with $\mathcal{\hat{R}}$ being time-reversal operator and $\mathcal{\hat{W}}$ being the bipartite lattice operator. 

Furthermore, we do not have to restrict ourselves to the dynamics of $S^z_i(t)$ and $n_i(t)$. For instance, if we consider the in-plane anti-ferromagnetic spin dynamics by measuring operator $(-1)^{i_x+i_y}\hat{c}^\dag_{i\uparrow}\hat{c}_{i\downarrow}$, because the particle-hole transformation $\mathcal{\hat{P}}$ maps this operator to the local pairing operator $\hat{c}^\dag_{i\uparrow}\hat{c}^\dag_{i\downarrow}$, the dynamics of the in-plane anti-ferromagnetic operator is therefore equivalent to the dynamics of the local pairing operator $\hat{c}^\dag_{i\uparrow}\hat{c}^\dag_{i\downarrow}$, under the same conditions as discussed above. Thus, we can formulate the most general version of the theorem as follows:

\textit{Theorem.}  For the FHM on a square lattice, the measurement of the operator $\hat{O}_1$ starting from a quantum state $|\Psi\rangle_1$ with interaction parameter $U_0$ and conserved quantities $N_\uparrow+N_\downarrow-N_\text{s}=x$ and $N_\uparrow-N_\downarrow=y$ is always equal to the measurement of the operator $\hat{O}_2$ starting from quantum state $|\Psi\rangle_2$ for the same interaction parameter $U_0$ and conserved quantities $N_\uparrow+N_\downarrow-N_\text{s}=y$ and $N_\uparrow-N_\downarrow=x$, provided that $\hat{O}_1$, $\hat{O}_2$, $|\Psi\rangle_1$ and $|\Psi\rangle_2$ satisfy the following conditions: 

(\textbf{1}) $|\Psi\rangle_1$ and $|\Psi\rangle_2$ are related by the particle-hole transformation $\mathcal{\hat{P}}$; and $\hat{O}_1$, $\hat{O}_2$ are also related by the particle-hole transformation $\mathcal{\hat{P}}$;

(\textbf{2}) Both $\hat{O}_2$ and $|\Psi\rangle_2$ are invariant under $\mathcal{\hat{S}} = \mathcal{\hat{R}}\mathcal{\hat{W}}$, with $\mathcal{\hat{R}}$ being time-reversal operator and $\mathcal{\hat{W}}$ being the bipartite lattice operator. 

Finally we would like to comment on the experimental relevance of this theorem. First of all, we should acknowledge that the initial states for either MIT experiment or the Princeton experiment is not the same spin-density-wave or the charge-density-wave state as we defined in Eq. \ref{SDW} or Eq. \ref{CDW}. Strictly speaking, our theorem does not rigorously apply. However, it is still worth checking whether this equivalence can hold approximately despite of the difference in the initial state. So far, as presented in Ref. \cite{MIT} and Ref. \cite{Princeton}, the MIT group has only reported spin transport measured for the half-filled FHM with zero spin imbalance, and the Princeton group has only reported data for charge transport of the FHM doped away from half-filling. Thus these two sets of data can not be directly compared with each other. However, it will be straightforward for them to extend their measurements to the regions with both finite doping and finite spin imbalance, and by comparing these data sets our theorem can be confirmed experimentally. On the other hand, with the Fermi gas microscope, it is also possible to prepare the state like Eq. \ref{SDW} and \ref{CDW} deterministically with single site addressing technique, as have been done for bosons \cite{Bloch}. In this way, our theorem can be directly confirmed experimentally. Our results establishes rigorously new relations of quantum dynamics in a highly non-equilibrium situation.   

\textit{Acknowledgment.} This work is supported MOST under Grant No. 2016YFA0301600 and NSFC Grant No. 11734010.

\end{document}